\documentclass[a4paper,11pt]{article}
\usepackage{latexsym}
\usepackage{amssymb}
\usepackage{amsmath}
\usepackage{authblk}
\title{Parameterized post Newtonian approximation in a teleparallel model of dark energy with a boundary term}

\author[1]{H. Mohseni Sadjadi\thanks{mohsenisad@ut.ac.ir}}

\affil[1]{Department of Physics, University of Tehran}

\begin{document}

\maketitle

\begin{abstract}

We study the parameterized post-Newtonian approximation in teleparallel model of gravity with a scalar field. The scalar field is  non-minimally coupled to the scalar torsion as well as
to the boundary term introduced in \cite{baha}.  We show that, in contrast to the case where the scalar field is only coupled to the scalar torsion, the presence of the new coupling affects
the parameterized post-Newtonian parameters. These parameters for different situations are obtained and discussed.
\end{abstract}

\section{Introduction}
In teleparallel model of gravity, instead of the torsionless Levi-Civita connections, curvatureless Weitzenb\"{o}ck  connections are used \cite{tele}. Teleparallel equivalent of general relativity was first introduced in \cite{ein} as an attempt for unification of electromagnetism and gravity. This theory is considered as an alternative theory of usual general relativity and has been recently employed to study the late time acceleration of the Universe \cite{acc}.  This can be accomplished by considering modified $f(T)$ models \cite{mod}, where $T$ is the torsion scalar, or by introducing exotic field such as quintessence. Assuming a nonminimal coupling between the scalar field and the torsion opens new windows in studying the cosmological evolution \cite{quint}, and can be viewed as a promising scenario for late time acceleration and super-acceleration \cite{sad}.

A non minimally coupled scalar field, like scalar-tensor model, may alter Newtonian potential. So it is necessary to check if the model can pass local gravitational tests such as solar system observations. This can be done in the context of the parameterized post-Newtonian formalism  \cite{ppn}. In  \cite{Wei}, and \cite{Li} it was shown that
when the scalar field is only coupled to the scalar torsion, there is no deviation from general relativity in the parameterized post-Newtonian (PPN) parameters and the theory is consistent with gravitational tests and solar system observations.

Recently a new coupling between the scalar field and a boundary term $\mathcal{B}$, corresponding to torsion divergence $\mathcal{B}\propto \nabla_\mu T^\mu$, was introduced in \cite{baha}, where the cosmological consequences of such a coupling for some simple power law scalar field potential, and the stability of the model were discussed. There was found that
that the system evolves to an attractor solution, corresponding to late time acceleration, without any fine tuning of the parameters. In this framework, the phantom divide line crossing is also possible. Thermodynamics aspects of this model were studied in \cite{thermo}. This model includes two important subclasses, i.e. non minimally coupled quintessence to the Ricci scalar and non minimally coupled quintessence to the scalar torsion. Another important feature of this model is its ability to describe the present cosmic acceleration in the framework of $Z_2$ symmetry breaking by alleviating the coincidence problem \cite{jcap}.

In this paper, we aim to investigate whether this new boundary coupling may affect the Newtonian potential and PPN parameters: $\gamma(r)$ and $\beta(r)$.

The scheme of the paper is as follows:  In the second section we introduce the model and obtain the equations of motion.  In the third section, we obtain the weak field expansion of the equations in the PPN formalism and obtain and discuss their solutions for spherically symmetric metric. We show that the PPN parameters may show deviation from general relativity. We consider different special cases and derive explicit solutions for the PPN parameters in terms of the model parameters and confront them with observational data.

We use units $\hbar=c=1$ and choose the signature $(-,+,+,+)$ for the metric.

\section{The model and the field equations}
In our study we use vierbeins $e_a={e_a}^\mu\partial_\mu$ whose duals, ${e^a}_\mu$, are defined through ${e^a}_\mu {e_a}^\nu=\delta^\nu_\mu$.  The metric tensor is given by $g^{\mu \nu}=\eta_{ab}{e_a}^\mu {e_b}^\nu$,
$\eta=diag(-1,1,1,1)$. $e=det({e^a}_\mu) =det\sqrt{-g}$. Greek indices (indicating coordinate bases) like the first Latin indices (indicating orthonormal bases) $a,b,c,..$ belongs to $\{0,1,2,3\}$, while $i,j,k,..\in \{1,2,3\}$.

Our model is specified by the action \cite{baha}:
\begin{equation}\label{1}
S=\int\left({T\over 2k^2}+{1\over 2}\left(-\partial_\mu\phi\partial^{\mu}\phi+\epsilon T \phi^2+\chi \mathcal{B}\phi^2\right)-V(\phi)+\mathcal{L}_m\right) ed^4x,
\end{equation}
where $k^2=8\pi G_N$, and  $G_N$ is Newtonian gravitational constant.
The torsion scalar is defined by
\begin{equation}\label{2}
T={S^\rho}_{\mu \nu}{T_\rho}^{\mu \nu}={1\over 4}{T^\rho}_{\mu \nu}{T_\rho}^{\mu \nu}
+{1\over 2}{T^\rho}_{\mu \nu}{T^{\nu \mu}}_{\rho}-{T^\rho}_{\mu \rho}{T^{\nu \mu}}_{\nu},
\end{equation}
and the boundary term is  \cite{boundary}
\begin{equation}\label{3}
\mathcal{B}={2\over e}\partial_\mu\left(eT^\mu\right),
\end{equation}
where $T^\mu={{T^\lambda}_{\lambda}}^\mu$. The  Weitzenb\"{o}ck torsion, and connection are given by
\begin{equation}\label{p1}
{T^\lambda}_{\mu \nu}={\Gamma^{\lambda}}_{\mu \nu}-{\Gamma^{\lambda}}_{\nu \mu}={e_a}^{\lambda}{T^a}_{\mu \nu},
\end{equation}
and
\begin{equation}\label{p2}
{\Gamma^\lambda}_{\mu \nu}={e_a}^\lambda \partial_\mu {e^a}_\nu,
\end{equation}
respectively.
${S^\rho}_{\mu \nu}$ is defined according to
\begin{equation}\label{p3}
{S^\rho}_{\mu \nu}={1\over 4}\left({T^\rho}_{\mu \nu}-{T_{\mu \nu}}^{\rho}+{T_{\nu \mu}}^\rho\right)+{1\over 2}\delta^\rho_\mu {T^{\sigma}}_{\nu \sigma}-{1\over 2}\delta_\nu ^\rho {T^\sigma}_{\mu \sigma}.
\end{equation}

Note that $R=-T+\mathcal{B}$, where $R$ is the Ricci scalar curvature. Hence for $\chi=-\epsilon$ the model reduces to a quintessence model coupled
non- minimally to the scalar curvature, while for $\chi=0$, we recover the quintessence model coupled non-minimally to the scalar torsion.

By variation of the action (\ref{1}) with respect to the vierbeins we obtain
\begin{eqnarray}\label{4}
&&\left({2\over k^2}+2\epsilon\phi^2\right)\left(e^{-1} {e^a}_\mu \partial_\lambda(e {S_a}^{\lambda \nu})-{T^\rho}_{\beta \mu}{S_\rho}^{\nu \beta}-{1\over 4} \delta^\nu_\mu T\right)\nonumber \\
&&-\delta^\nu_\mu\left(-{1\over 2}\partial_\alpha \phi \partial^\alpha \phi -V(\phi)\right)-\partial^{\nu}\phi\partial_\mu\phi+4(\chi +\epsilon)\phi{S_\mu}^{\beta \nu} \partial_\beta \phi\nonumber \\
&&+\chi\left(\delta_\mu^\nu \Box \phi^2-\nabla^\nu \nabla_\mu \phi^2\right)=-\tau^\nu_\mu.
\end{eqnarray}
$\tau^\nu_\mu$ is the energy momentum tensor of matter.

The trace of (\ref{4}), multiplied by  $-\delta^\nu_\mu/2$,  is

\begin{eqnarray}\label{11}
 &&-\delta^\nu_\mu \left({1\over k^2}+\epsilon \phi^2\right)\left(e^{-1}{e^a}_\alpha \partial_\lambda(e{S_a}^{\lambda \alpha})\right)-{1\over 2}\delta^\nu_\mu\partial_\alpha \phi\partial^\alpha \phi-2\delta^\nu_\mu V(\phi)\nonumber \\
&& -2\delta^\nu_\mu(\chi+\epsilon)\phi {S_\alpha}^{\beta \alpha}\partial_\beta \phi-{3\over 2}\chi\delta^{\nu}_\mu\Box \phi^2={1\over 2}\delta_\mu^\nu \tau.
\end{eqnarray}
By combining (\ref{11}) and (\ref{4}) we obtain
\begin{eqnarray}\label{12}
&&\left({2\over k^2}+2\epsilon \phi^2\right)\left( e^{-1}{e^a}_\mu\partial_\lambda(e{S_a}^{\lambda \nu})-{T^\rho}_{\beta \mu}{S_\rho}^{\nu \beta}-{1\over 4}\delta_\mu^\nu T\right)-\delta_\mu^\nu V(\phi)\nonumber \\
&&-\partial^\nu \phi\partial_\mu \phi+4(\chi+\epsilon) \phi{S_\mu}^{\beta \nu}\partial_\beta \phi-\chi\nabla^\nu \nabla_\mu \phi^2-{1\over 2}\chi\delta^\nu_\mu \Box \phi^2\nonumber \\
&&-\delta^\nu_\mu\left({1\over k^2}+\epsilon \phi^2\right)\left(e^{-1}{e^a}_\alpha\partial_\lambda e{S_a}^{\lambda \alpha}\right)-2\delta^\nu_\mu (\chi+\epsilon)\phi {S_\alpha}^{\beta \alpha}\partial_\beta \phi\nonumber \\
&&=-\tau^\nu_\mu+{1\over 2}\delta^\nu_\mu \tau.
\end{eqnarray}
Note that  the trace of the energy momentum tensor is $\tau=g^{\mu \nu}\tau_{\mu \nu}$.

In the same way, variation of the action with respect the scalar field gives
\begin{equation}\label{13}
-{1\over e}\partial_\mu e g^{\mu \nu}\partial_\nu \phi-\chi \mathcal{B} \phi-\epsilon T \phi+V'(\phi)=0.
\end{equation}
Eqs. (\ref{12}) and (\ref{13}) are the main equations that we will work with in the following parts.

\section{post-Newtonian formalism}
To investigate the post-Newtonian approximation \cite{ppn} of the model, the perturbation is specified by the velocity of the source matter $\left|\vec{v}\right|$ such that e.g.  $\mathcal{O}(n)\sim \left|\vec{v}\right|^n$. The matter source is assumed to be a perfect fluid obeying the post-Newtonian hydrodynamics:
\begin{equation}\label{5}
\tau_{\mu \nu}=(\rho+\rho \Pi+p) u_{\mu}u_{\nu}+pg_{\mu \nu},
\end{equation}
where $\rho$ is energy density, $p$ is the pressure and $\Pi$ is the specific internal energy. $u^\mu$ is the four-vector velocity of the fluid. The velocity of the source matter is $v^i={u^i\over u^0}$. The orders of smallness of energy momentum tensor ingredients are \cite{ppn}
\begin{equation}\label{p3}
\rho\sim \Pi \sim {p\over \rho}\sim U \sim \mathcal{O}(2)
\end{equation}
where $U$ is the Newtonian gravitational potential. The components of the energy momentum tensor are given by
\begin{eqnarray}\label{6}
&&{\tau_0}^0=-\rho-\rho v^2-\rho \Pi+\mathcal{O}(6)\nonumber \\
&&{\tau_0}^i=-\rho v^i +\mathcal{O}(5)\nonumber \\
&&{\tau_i}^j=\rho v^j v_i+p\delta_i^j+\mathcal{O}(6).
\end{eqnarray}

We expand the metric around the Minkowski flat background as \cite{Li,Wei}
\begin{equation}\label{7}
g_{\mu \nu}=\eta_{\mu \nu}+{h^{(2)}}_{\mu \nu}+{h^{(3)}}_{\mu \nu}+{h^{(4)}}_{\mu \nu}+\mathcal{O}(5)
\end{equation}
Note ${h^{(1)}}_{\mu \nu}=0$ \cite{ppn}. Accordingly, the vierbeins may be expanded as \cite{Li}
\begin{equation}\label{8}
{e^a}_\mu=\delta^a _\mu+ {{B^{(2)}}^a}_\mu+{{B^{(3)}}^a}_\mu+{{B^{(4)}}^a}_\mu+\mathcal{O}(5),
\end{equation}
Note ${{B^{(1)}}^a}_\mu=0$.  In our analysis we need non zero components of the metric up to order 4,  i.e  : $h^{(2)}_{ij},\,\, h^{(2)}_{00},\,\,  h^{(3)}_{0i},\,\, h^{(4)}_{00}$. We also use the notation
$B_{\mu \nu}=\eta_{\mu \sigma}{B^\sigma}_\nu$ and ${\delta_a}^\sigma {B^a}_\nu={B^\sigma}_\nu$. By comparing (\ref{7}) and (\ref{8}) we derive (like \cite{Li}, and \cite{Wei}, $B^{(2)}_{ij}$ is assumed to be diagonal)
\begin{eqnarray}\label{9}
&&h^{(2)}_{ij}=2B^{(2)}_{ij}\nonumber \\
&&h^{(2)}_{00}=2B^{(2)}_{00}\nonumber \\
&&h^{(3)}_{0i}=2B^{(3)}_{0i}\nonumber \\
&&h^{(4)}_{00}=2B^{(4)}_{00}-(B^{(2)}_{00})^2.
\end{eqnarray}
We introduce two functions $A$, and $\gamma$ (which is one of the PPN parameters) through \cite{Wei}
\begin{eqnarray}\label{10}
&&B^{(2)}_{00}=A\nonumber \\
&&B^{(2)}_{ij}=\gamma A \delta_{ij}.
\end{eqnarray}

The scalar field is expanded as
\begin{equation}\label{14}
\phi=\phi_0+\psi,
\end{equation}
where
\begin{equation}\label{15}
\psi=\psi^{(2)}+\psi^{(4)}+\mathcal{O}(6),
\end{equation}
and $\phi_0$ is a constant cosmological background. $\phi_0$ is of order $\mathcal{O}(0)$ and may evolve in times of order of the Hubble time, so in solar system tests we assume that it is static. The time derivative, $\partial_0={\partial\over \partial t}$, of the other fields are weighted with order $\mathcal{O}(1)$ \cite{ppn}.

The potential around the background is
\begin{equation}\label{16}
V(\phi)=V(\phi_0)+V'(\phi_0)\psi+{V''(\phi_0)\over 2} \psi^2+\mathcal {O}(6).
\end{equation}

Defining $V(\phi_0)=V_0,\,\, {V^{(n)}(\phi_0)\over n!}=V_n$ we find
\begin{equation}\label{17}
V'=V_1+2V_2\psi+3V_3\psi^2+\mathcal{O}(6).
\end{equation}

After these preliminaries, let us solve the eqs. (\ref{12}) and (\ref{13}) order by order in the PPN formalism.
At zeroth order (\ref{12}) and (\ref{13}) imply
\begin{equation}\label{18}
V_0=V_1=0.
\end{equation}

0-0 component of (\ref{12}) gives
\begin{eqnarray}\label{19}
&&\left({2\over k^2}+2\epsilon \phi^2\right)\left( e^{-1}{e^a}_0\partial_\lambda(e{S_a}^{\lambda 0})-{T^\rho}_{\beta 0}{S_\rho}^{0 \beta}-{1\over 4}T\right)-V(\phi)-\partial^0 \phi\partial_0 \phi\nonumber \\
&&+4(\chi+\epsilon) \phi{S_0}^{\beta 0}\partial_\beta \phi-\chi\nabla^0\nabla_0 \phi^2-{1\over 2}\chi\Box \phi^2-\left({1\over k^2}+\epsilon \phi^2\right)\left(e^{-1}{e^a}_\alpha\partial_\lambda e{S_a}^{\lambda \alpha}\right)\nonumber \\
&&-2(\chi+\epsilon)\phi {S_\alpha}^{j \alpha}\partial_j \phi=-\tau^0_0+{1\over 2}\tau,
\end{eqnarray}
which at order 2 reduces to:
\begin{equation}\label{20}
\left({1\over k^2}+\epsilon \phi_0^2\right)\partial_j{S_0}^{j 0}-V(\phi)-{1\over 2}\chi\Box \phi^2-\left({1\over k^2}+\epsilon \phi^2\right)\partial_j {S_i}^{j i}={\rho\over 2},
\end{equation}
resulting in
\begin{equation}\label{21}
-\left({1\over k^2}+\epsilon \phi_0^2\right)\nabla^2 A-\chi \phi_0\nabla^2 \psi^{(2)}=-{1\over k^2}\nabla^2 U,
\end{equation}
where the potential is given by
\begin{equation}\label{25}
\nabla^2U=-{k^2\over 2}\rho.
\end{equation}
To obtain (\ref{21}), we have used
\begin{eqnarray}\label{22}
&&{{S^{(2)}}^0}_{j0}=-\partial_j(\gamma A),\,\,\,{{S^{(2)}}^j}_{ij}=\partial_i\left((1-\gamma)A\right),\,\,\,{{S^{(2)}}^i}_{0j}=0\nonumber \\
&&\partial_\mu e^{(2)}=\partial_\mu \left((3\gamma -1)A\right),\,\,\, {{T^{(2)}}^{0}}_{i0}=-\partial_i A,\,\,\,{{S^{(2)}}^0}_{0i}=\partial_i(\gamma A).
\end{eqnarray}
By taking the trace of i-j component of (\ref{12}), at order 2, we obtain:
\begin{equation}\label{23}
-3\left({1\over k^2}+\epsilon\phi_0^2\right)\partial_j{S_0}^{j0}-\left({1\over k^2}+\epsilon \phi_0^2\right)\partial_j{S_i}^{ji}-5\chi\phi_0\nabla^2\psi^{(2)}=-{3\over 2}\rho,
\end{equation}
which reduces to
\begin{equation}\label{24}
\left({1\over k^2}+\epsilon\phi_0^2\right)\nabla^2\left((4\gamma -1)A\right)-5\chi\phi_0\nabla^2\psi^{(2)}={3\over k^2}\nabla^2U.
\end{equation}

At the second order perturbation, the boundary term $\mathcal{B}$, defined in (\ref{3}), is derived as
\begin{equation}\label{26}
\mathcal{B}^{(2)}=2\nabla^2\left((1-2\gamma)A\right).
\end{equation}
Hence from (\ref{13}) the equation of motion of the scalar field becomes
\begin{equation}\label{27}
-\nabla^2 \psi^{(2)}+2V_2 \psi^{(2)}=2\chi (1-2\gamma)A \phi_0.
\end{equation}
Eqs. (\ref{21}), (\ref{24}), and (\ref{27})  are our three main equations for determining $A$, $\gamma$, and $\psi^{(2)}$.
Using these three equations, for a given $U$, $A$ is derived as
\begin{equation}\label{28}
A={2\over (1+\epsilon\phi_0^2 k^2)(1+\gamma)}U,
\end{equation}
and $\psi^{(2)}$ is obtained as
\begin{equation}\label{29}
\psi^{(2)}={\gamma -1\over k^2\chi \phi_0(\gamma+1)}U.
\end{equation}
$\gamma$ is determined by the equation
\begin{equation}\label{30}
\left(1-{6k^2\chi^2\phi_0^2\over 1 +\epsilon k^2 \phi_0^2}\right)\nabla^2(\Gamma U)-2V_2(\Gamma U)=-{k^4\chi^2\phi_0^2\over 1+\epsilon k^2\phi_0^2}\rho,
\end{equation}
where $\Gamma:={\gamma-1\over \gamma+1}$.
(\ref{30}) is a nonhomogeneous screened Poisson equation whose solution is
\begin{equation}\label{31}
\Gamma U={k^4 \chi^2 \phi_0^2\over  1+\epsilon k^2\phi_0^2-6k^2\chi^2\phi_0^2}\int{\exp{\left(-\lambda \left|\vec{r}-\vec{r'}\right|\right)}\over 4\pi\left|\vec{r}-\vec{r'}\right|}\rho(x',t) d^3x',
\end{equation}
where
\begin{equation}\label{55}
\lambda=\sqrt{{2V_2(1+k^2 \epsilon \phi_0^2)\over 1 +k^2 \epsilon \phi_0^2-6 k^2 \chi^2 \phi_0^2}}.
\end{equation}
Equation (\ref{28}) allows us to take
\begin{equation}\label{32}
G={2\over (1+k^2 \epsilon \phi_0^2)(\gamma+1)},
\end{equation}
where $G$
is defined through
\begin{equation}\label{33}
h_{00}^{(2)}=2A=2GU.
\end{equation}
So one can define an effective $G_{eff.}$ through
\begin{equation}
G_{eff}=G G_N
\end{equation}

$0-i$ component of (\ref{4}) at the third order gives
 \begin{equation}\label{35}
\left({2\over k^2}+2\epsilon \phi_0^2\right)\partial_\mu{S_0}^{\mu i}=-{{\tau^{(3)}}_0}^i=\rho v^i,
\end{equation}
which by using
\begin{eqnarray}\label{34}
 &&{{T^{(3)}}^0} _{ij}=\partial_i{{B^{(3)}}^0}_j-\partial_j {{B^{(3)}}^0}_i\nonumber\\
&& {{T^{(3)}}^i} _{j0}=\partial_j{{B^{(3)}}^i}_0-\delta^i_j\partial_0(\gamma A)\nonumber\\
 &&{{T^{(3)}}^i}_{i0}=-{3}\partial_0(\gamma A)+3\partial_i {{B^{(3)}}^i}_0,
\end{eqnarray}
reduces to
\begin{equation}\label{36}
\left({2\over k^2}+2\epsilon \phi_0^2\right)\left(\partial_0\partial_i (\gamma A)-{1\over 2}\nabla^2{{B^{(3)}}^0}_i+{1\over 2}\partial^j\partial_i {{B^{(3)}}^0}_j\right)=\rho v_i.
\end{equation}

To simplify computations one may employ gauge condition

\begin{eqnarray}\label{38}
&& -\partial^j{{B^{(2)}}^i}_j+{1\over 2}\partial^i {{B^{(2)}}^\mu}_\mu={\chi k^2\phi_0\over k^2+\epsilon \phi_0^2}\partial^i \psi^{(2)}\nonumber \\
&&-\partial_j{{B^{(3)}}^j}_0+{1\over 2}\partial_0 {{B^{(2)}}^j}_j={\chi k^2\phi_0\over k^2+\epsilon \phi_0^2}\partial_0 \psi^{(2)},
\end{eqnarray}
which determines ${{B^{(3)}}_0}^j$ in terms of second order parameters. This gauge is compatible with eqs. (\ref{21}) and (\ref{24}).

Using
\begin{eqnarray}\label{39}
&&{{S^{(4)}}^i}_{ji}=\gamma A\partial_j(\gamma A)+A\partial_j A-\partial_j {B^{(4)}}_0^0+\partial_0{{B^{(3)}}^0}_j\nonumber \\
&&{{S^{(4)}}^0}_{j0}=\gamma A\partial_j(\gamma A)\nonumber \\
&&{{S^{(3)}}_{i0}}^i=-{3\over 2}\partial_0(\gamma A),
\end{eqnarray}
one can find that (\ref{19}) at the fourth order gives
\begin{eqnarray}\label{40}
&&\left({1\over k^2}+\epsilon \phi_0^2\right)\left(\nabla^2{{B^{(4)}}^0}_0+\nabla^2(\gamma A)^2-3\nabla(\gamma A).\nabla A-A\nabla^2 A\right)\nonumber \\
&&-4\epsilon\phi_0\psi^{(2)}\nabla^2 A-2(\chi +\epsilon)\phi_0\nabla \psi^{(2)}.\nabla A-{\chi\over 2}\nabla^2 (\psi^{(2)})^2-V_2(\psi^{(2)})^2\nonumber \\
&&-\chi \phi_0\nabla^2 \psi^{(4)}+3\chi \phi_0 \nabla((\gamma-1)A).\nabla\psi^{(2)}+\left(\partial_0\psi^{(2)}\right)^2\nonumber \\
&&+3\chi \phi_0 \partial_0^2\psi^{(2)}+\left({1\over k^2}+\epsilon \phi_0^2\right)\partial_0\left(3\partial_0 (\gamma A)-\partial_i {{B^{(3)}}^i}_0\right)\nonumber \\
&&-\left({1\over k^2}+\epsilon \phi_0^2\right)\partial^j\partial_0{{B^{(3)}}^0}_j={1\over 2}\tau^{(4)}-{\tau^{(4)}}^0_0.
\end{eqnarray}
Also, the scalar field equation at the fourth order is
\begin{equation}\label{41}
-\nabla^2 \psi^{(4)}+2V_2\psi^{(4)}+\partial_0^2\psi^{(2)}=\chi \phi_0 B^{(4)}+\psi^{(2)} B^{(2)}+\epsilon \phi_0 T^{(4)}-3 V_3 (\psi^{(2)})^2.
\end{equation}
By using
\begin{eqnarray}\label{42}
\mathcal{B}^{(4)}&=&-8\nabla^2\left(\gamma^2 A^2\right)+14\nabla.\left(\gamma A\nabla A\right)+2(1-5\gamma)A\nabla^2A\nonumber \\
&+&12\gamma A\nabla^2(\gamma A)-\nabla^2{{B^{(4)}}^0}_0+6\partial_0^2(\gamma A)-2\partial_i\partial_0{{B^{(3)}}^i}_0,
\end{eqnarray}
and
\begin{equation}\label{43}
T^{(4)}=2\nabla(\gamma A).\nabla\left((2-\gamma)A\right),
\end{equation}
(\ref{41}) becomes
\begin{eqnarray}\label{44}
&&-\nabla^2 \psi^{(4)}+2V_2\psi^{(4)}+\partial_0^2\psi^{(2)}=6\chi\phi_0\partial_0^2(\gamma A)-2\chi \phi_0\partial_i\partial_0 {{B^{(3)}}^i}_0\nonumber \\
&&2\chi \phi_0\left(-4\gamma^2 A^2+7\nabla.(\gamma A\nabla A)+(1-5 \gamma)A\nabla^2 A+6\gamma A\nabla^2(\gamma A)\right)\nonumber \\
&&+2\psi^{(2)}\nabla^2((1-2\gamma)A)+2\epsilon \phi_0 \nabla(\gamma A).\nabla((2-\gamma)A)-3V_3(\psi^{(2)})^2\nonumber \\
&&-2\chi \phi_0\nabla^{2}{{B^{(4)}}^0}_0+3\nabla(1-\gamma)A.\nabla \psi^{(2)}.
\end{eqnarray}

(\ref{40}), and (\ref{44}), are our main results in the fourth order. These equations together with (\ref{36}) and (\ref{38}) in the third order, and (\ref{28}), (\ref{29}), (\ref{30}), in the second order must be solved to give the post-Newtonian parameters.

To solve these complicated equations, we consider solutions specified by
$U=U(r)$ which results in
\begin{equation}\label{45}
A=A(r),\,\,\,\gamma=\gamma(r),\,\, \psi^{(2)}=\psi^{(2)}(r).
\end{equation}
The gauge (\ref{38}) implies $\partial^j{{B^{(3)}}^0}_j=0$. Therefore (\ref{36}) reduces to
\begin{equation}\label{46}
-\left({1\over k^2}+\epsilon\phi_0^2\right)\left(\nabla^2{{B^{(3)}}^0}_i\right)=\rho v_i.
\end{equation}
For $v^i=0$, (\ref{46}) gives ${{B^{(3)}}^0}_i=0$ (by assumption that perturbation terms vanish at large distance).
In this situation eqs. (\ref{40}) and (\ref{44}) become
\begin{eqnarray}\label{47}
&&\left({1\over k^2}+\epsilon \phi_0^2\right)\left(\nabla^2{{B^{(4)}}^0}_0+\nabla^2(\gamma A)^2-3\nabla(\gamma A).\nabla A-A\nabla^2 A\right)\nonumber \\
&&-4\epsilon\phi_0\psi^{(2)}\nabla^2 A-2(\chi +\epsilon)\phi_0\nabla \psi^{(2)}.\nabla A-{\chi\over 2}\nabla^2 (\psi^{(2)})^2-V_2(\psi^{(2)})^2\nonumber \\
&&-\chi \phi_0\nabla^2 \psi^{(4)}+3\chi \phi_0 \nabla((\gamma-1)A).\nabla\psi^{(2)}\nonumber \\
&&={1\over 2}\tau^{(4)}-{\tau^{(4)}}^0_0,
\end{eqnarray}
and
\begin{eqnarray}\label{48}
&&-\nabla^2 \psi^{(4)}+2V_2\psi^{(4)}=\nonumber \\
&&2\chi \phi_0\left(-4\gamma^2 A^2+7\nabla.(\gamma A\nabla A)+(1-5 \gamma)A\nabla^2 A+6\gamma A\nabla^2(\gamma A)\right)\nonumber \\
&&+2\psi^{(2)}\nabla^2((1-2\gamma)A)+2\epsilon \phi_0 \nabla(\gamma A).\nabla((2-\gamma)A)-3V_3(\psi^{(2)})^2\nonumber \\
&&-2\chi \phi_0\nabla^{2}{B^{(4)}}_0^0+3\nabla(1-\gamma)A.\nabla \psi^{(2)},
\end{eqnarray}
respectively.
To obtain post-Newtonian parameters we must obtain $A$, $\psi^{(2)}$, and $\gamma(r)$. By inserting them in (\ref{47}) and (\ref{48}), we obtain solutions for ${B^{(4)}}_0^0$.
To do so we consider a spherically symmetric metric with a point source.

\subsection{Spherically symmetric metric }
The source is assumed to be
\begin{equation}\label{49}
\rho=M\delta(\vec{r}),\,\,\,\Pi=0,\,\,\, p=0,\,\,\, v_i=0,
\end{equation}
and the metric is given by
\begin{eqnarray}\label{50}
&&g_{00}=-1+2G_{eff}U-2G_{eff}^2\beta U^2+ Self +\mathcal{O}(6)\nonumber \\
&&g_{ij}=\mathcal{O}(5)\nonumber \\
&&g_{ij}=\left(1+2G_{eff}\gamma U \right)\delta_{ij}+\mathcal{O}(4),
\end{eqnarray}
where "Self" denotes self-energy terms of order 4, and $\beta$ is the PPN parameter.  The Newtonian potential is
\begin{equation}\label{u}
U={k^2M\over 8\pi r}.
\end{equation}
To determine $\gamma$,
from (\ref{28}), (\ref{29}), and (\ref{31}),  we obtain
\begin{equation}\label{51}
\psi^{(2)}={2\chi \phi_0\over 1+\epsilon k^2\phi_0^2-6\chi^2k^2\phi_0^2 }\exp(-\lambda r),
\end{equation}
and
\begin{equation}\label{52}
A={k^2 M\over 4\pi (1+\epsilon k^2\phi_0^2)(1+\gamma)r},
\end{equation}
where
\begin{equation}\label{53}
\gamma={1+\alpha \exp(-\lambda r)\over 1-\alpha \exp(-\lambda r)},
\end{equation}
 in which
\begin{equation}\label{54}
\alpha={2 k^2 \chi^2\phi_0^2\over 1+k^2\epsilon \phi_0^2 -6 k^2 \chi^2 \phi_0^2},
\end{equation}
and $\lambda$ is given by (\ref{55}). From  $h_{00}^{(2)}=2A=2GU$, we obtain $G$ as (\ref{32}).

To obtain ${B^{(4)}}_0^0$, one must insert (\ref{51}), (\ref{52}) and (\ref{53} )in (\ref{47}) and (\ref{48}), and solve them together.  From  ${B^{(4)}}_0^0$ we determine the other PPN parameter, $\beta$, as
\begin{equation}\label{57}
2{{B^{(4)}}^0}_0+A^2=2G^2\beta(r)U^2(r).
\end{equation}

To determine PPN parameters, $\gamma$ and $\beta$,  we will consider different situations.

\subsubsection{$\chi=0$}
for $\chi=0$, from (\ref{53}) and (\ref{54}),  we find $\gamma=1$, hence
\begin{equation}\label{A1}
A={k^2 M\over 8\pi (1+\epsilon k^2\phi_0^2)r},\,\,\,\, G={1\over 1+k^2 \epsilon \phi_0^2}.
\end{equation}
(\ref{51}) gives $\psi^{(2)}=0$.  So we write (\ref{47}) as
\begin{equation}\label{59}
\nabla^2{{B^{(4)}}^0}_0-{1\over 2}\nabla^2(A)^2+2A\nabla^2 A=0,
\end{equation}
where $\nabla A.\nabla A={1\over 2}\nabla^2 A^2-A\nabla^2A$ has been used. Putting (\ref{A1}) in (\ref{59}), and ignoring gravitational self-energy, we obtain
\begin{equation}\label{58}
{{B^{(4)}}^0}_0=-{A^2\over 2}+{k^4 M^2\over 64\pi^2(1+\epsilon k^2\phi_0^2)^2 r^2}.
\end{equation}
Therefore (\ref{57}) yields
$\beta(r)=1$. So for $\chi=0$ we find
\begin{equation}\label{60}
\beta(\chi=0)=\gamma(\chi=0)=1
\end{equation}
Therefore there is no deviation from general relativity for the PPN parameters. This is in complete agreement with \cite{Li} and \cite{Wei}.

\subsection{$\phi_0=0$}
For $\chi\neq 0$, we may have also a situation with no deviation in PPN parameters from general relativity, this occurs for $\phi_0=0$. For example for potentials
\begin{equation}
V(\phi)=-{1\over 2} \mu^2 \phi^2+{\Lambda\over 4}\phi^4,\,\,\, \Lambda>0,
\end{equation}
and
\begin{equation}
V(\phi)=\Lambda \phi^n,\,\,\, \Lambda>0,\,\,\, n>1,
\end{equation}
$V_0=V_1=0$ (see (\ref{18})) leads to  $\phi_0=0$ which by using (\ref{53}-\ref{57}) results in $\gamma=1$, $G=1$, and $\beta=1$.
Therefore in this case too, there is no deviation from general relativity for the PPN parameters.

\subsubsection{$V(\phi)=0$}
If we ignore the scalar field potential, we obtain $\lambda=0$ (see (\ref{55})), and $\gamma$ becomes a constant
\begin{equation}\label{g1}
\gamma={1-(4\chi^2-\epsilon)k^2\phi_0^2\over 1-(8\chi^2-\epsilon)k^2\phi_0^2}.
\end{equation}
By solving the system of equations (\ref{47}) and (\ref{48}) for ${{B^{(4)}}^0}_0$  and by considering eqs. (\ref{51}-\ref{57}), after some computations we find
\begin{equation}\label{61}
\beta={P\over \left(1+(2\chi^2+\epsilon)k^2\phi_0^2\right)\left(1-(8\chi^2-\epsilon)k^2\phi_0^2\right)^2},
\end{equation}
where
\begin{eqnarray}\label{62}
&&P=1+160\big(\chi^6+{3\over 10}\epsilon \chi^5+{3\over 40}\epsilon \chi^4-{3\over 16}\epsilon^2\chi^3-{1\over 10}\epsilon^2\chi^2+\nonumber \\
&&{3\over 160}\epsilon^3\chi+{1\over 160}\epsilon^3\big)k^6\phi_0^6+2(\chi^3-8\chi^2+{3\over 2}\chi \epsilon+{3\over 2}\epsilon)k^2\phi_0^2+\nonumber \\
&&12(\chi^4-{7\over 3}\epsilon \chi^3-{8\over 3}\epsilon \chi^2+{1\over 2}\chi \epsilon^2+{1\over 4}\epsilon^2)k^4\phi_0^4.
\end{eqnarray}

Let us consider some limiting values:
For small $\chi$, $\chi\ll 1$ we have
\begin{eqnarray}\label{63}
\beta&=&1+{3\epsilon k^2\phi_0^2\over 1+\epsilon k^2\phi_0^2}\chi-{2 k^2\phi_0^2\over 1+\epsilon k^2\phi_0^2}\chi^2+\mathcal{O}(\chi^3)\nonumber \\
\gamma&=&1+{4k^2\phi_0^2\over 1+\epsilon k^2\phi_0^2}\chi^2+\mathcal{O}(\chi^4),
\end{eqnarray}
 and for small $k \phi_0$, $k\phi_0\ll 1$ we have
\begin{eqnarray}\label{64}
\beta&=&1+\chi(2\chi^2-2\chi+3\epsilon)k^2\phi_0^2+\mathcal{O}(k^4\phi_0^4)\nonumber \\
\gamma&=&1+4\chi^2 k^2\phi_0^2+\mathcal O(k^4\phi_0^4).
\end{eqnarray}

\subsubsection{$\lambda r\gg 1$}

In this limit from (\ref{51}) and (\ref{53}) we have $\psi^{(2)}=0$ and $\gamma=1$ respectively. The solution of (\ref{47}) is obtained as
\begin{equation}\label{65}
{{B^{(4)}}^0}_0={1\over 2}A^2+{\Omega+1\over 2\chi \phi_0}\psi^{(4)},
\end{equation}
where $\Omega=-1+{2\chi^2 k^2\phi_0^2\over 1+\epsilon k^2 \phi_0^2}$.
The equation of motion of the scalar field (\ref{48}) becomes
\begin{equation}\label{66}
\Omega \nabla^2 \psi^{(4)}+2V_2\psi^{(4)}=(\epsilon -2\chi)\phi_0\nabla^2 A^2+(8\chi -\epsilon)\phi_0A\nabla^2A,
\end{equation}
 whose solution, in the limit $\left|{V_2r\over \Omega}\right|\gg 1$, is
 \begin{equation}\label{67}
 \psi^{(4)}=\left({k^4M^2\phi_0(\epsilon-2\chi)\over 64 \pi^2(1+\epsilon k^2\phi_0^2)(2\chi^2 k^2\phi_0^2-\epsilon k^2\phi_0^2-1)}\right){1\over r^2}.
 \end{equation}
 From (\ref{67}), (\ref{65}), and (\ref{57}), we find
\begin{equation}\label{68}
\beta={\epsilon (\chi -1)k^2\phi_0^2-1\over (2\chi^2-\epsilon)k^2\phi_0^2-1}.
\end{equation}

For small $k^2\phi_0^2$, $k^2\phi_0^2\ll 1$ this gives
\begin{equation}\label{69}
\beta= 1+(2\chi^2 -\chi \epsilon)k^2\phi_0^2+\mathcal{O}(k^4\phi_0^4),
\end{equation}
and for small $\chi$,  $\chi\ll 1$ gives
\begin{equation}\label{70}
\beta=1-{\epsilon k^2\phi_0^2\over 1+\epsilon k^2\phi_0^2}\chi+2{k^2\phi_0^2\over 1+\epsilon k^2 \phi_0^2}\chi^2+\mathcal{O}(\chi^3).
\end{equation}

Finally let us note that for small $\lambda r$, $\lambda r\ll 1$, we take  $\exp(-\lambda r)\simeq 1$. In this case $\gamma$ and $\beta$ take the same form as (\ref{g1}) and (\ref{61}) respectively.
\subsection{Range of parameters}

The most precise experimental value for $\gamma$ has been obtained from Cassini \cite{cas}. The bound on this parameter is \cite{living}
\begin{equation}\label{test1}
\left|\gamma-1\right|\lesssim  2.3\times10^{-5}.
\end{equation}
In this experiment the gravitational interaction, in terms of Astronomical Unit takes place at $r\simeq 7.44\times 10^{-3}AU$ \cite{jarv} .

The parameter $\beta$ is determined by lunar laser ranging experiments via the Nordtvedt effect \cite{lunar}. This test indicates the bound \cite{living}
\begin{equation}\label{test2}
\left|\beta-1\right|\lesssim 2.3 \times 10^{-4},
\end{equation}
at a gravitational interaction distance  $r=1AU$ \cite{jarv}. (\ref{test1}) and (\ref{test2}) restrict the parameters of our model.

For $V(\phi)=0$ , (\ref{test1})and (\ref{g1}) give
\begin{equation}
\left|{4k^2\chi^2\phi_0^2\over 1-(8\chi^2-\epsilon)k^2\phi_0^2}\right|\lesssim  2.3\times10^{-5}.
\end{equation}

In the limiting cases (\ref{63}) and (\ref{64}) we find
\begin{eqnarray}
&&\left|{4k^2\phi_0^2\over 1+\epsilon k^2\phi_0^2}\chi^2\right|\lesssim  2.3\times10^{-5}\nonumber \\
&&\left|{3\epsilon k^2\phi_0^2\over 1+\epsilon k^2\phi_0^2}\chi^2\right|\lesssim 2.3 \times 10^{-4},
\end{eqnarray}
and
\begin{eqnarray}
&&4\chi^2k^2\phi_0^2\lesssim  2.3\times10^{-5} \nonumber \\
&&\left|\chi(2\chi^2-2\chi+3\epsilon)k^2 \phi_0^2\right|\lesssim 2.3 \times 10^{-4},
\end{eqnarray}
respectively.

For $\lambda r\gg 1$, we have
\begin{equation}
\sqrt{{2V_2(1+k^2 \epsilon \phi_0^2)\over 1 +k^2 \epsilon \phi_0^2-6 k^2 \chi^2 \phi_0^2}}\gg (1AU)^{-1},
\end{equation}
and (\ref{test2}) restricts our parameters as
\begin{equation}
\left|{\chi(2\chi-\epsilon)k^2\phi_0^2\over (2\chi^2-\epsilon)k^2\phi_0^2-1}\right|\lesssim 2.3 \times 10^{-4}.
\end{equation}

\section{conclusion}

The teleparallel model of gravity with a quintessence (nonminimally) coupled to the torsion and also to a boundary term (proportional to the torsion divergence) was considered (see (\ref{1})).  Although the model shows some interesting aspects in cosmology and in describing the late time acceleration of the Universe, but it must also pass local gravitational and solar system tests. So we studied the parameterized post-Newtonian (PPN) approximation of the model. We obtained the equations of motion (see the second section), and solve them order by order to obtain PPN parameters (see the third section).
Explicit expression for PPN parameters in spherically symmetric metric were obtained and different possible situations were discussed. Our results show that PPN parameters, except for some special cases, i.e. in the absence of boundary term and also zero scalar field background, differ from general relativity. So we conclude that, coupling of the scalar field to the  boundary term generally makes the model deviate from general relativity in the PPN limit.

 Since $T$ and $\mathcal{B}$ are not invariant under local Lorentz transformations, the teleparallel model with the boundary term is not invariant under Lorentz transformations unless one takes $\chi=-\epsilon$. Despite this, in spacetimes with spherical symmetry like Schwarzschild spacetime and so on, it is possible to choose good or preferred tetrads to solve this issue \cite{tam}. In scalar-tetrad theories of gravity the preferred tetrads cannot be detected via measuring the metric components \cite{hay}. Similarly, in our model, PPN parameters in the standard post-Newtonian formalism do not identify the effect of preferred tetrads. To include these effects one must generalize the post-Newtonian approach, as was pointed out in \cite{Wei}.


\begin{thebibliography}{99}

\bibitem{baha}S. Bahamonde and M. Wright, Phys. Rev. D 92, 084034 (2015), arXiv:1508.06580v4 [gr-qc]

\bibitem{tele}A. Unzicker and T. Case, arXiv:physics/0503046 [physics.hist-ph]; K.
Hayashi and T. Nakano, Prog. Theor. Phys. 38, 491 (1967); C. Pellegrini, J. Plebanski, Mat. Fys. Skr. Dan. Vid. Selsk. 2 (4), 1–39 (1963)
\bibitem{ein}A. Einstein, Sitzber. Preuss. Akad. Wiss. 217- 221 (1928) 
\bibitem{acc}E. V. Linder, Phys. Rev. D 81, 127301 (2010); H. Wei, Phys. Lett. B 712, 430 (2012); H. M. Sadjadi, Phys. Rev. D 92, 123538 (2015)
\bibitem{mod}Y.F. Cai, S. Capozziello, M. De Laurentis, E.N. Saridakis, Rept. Prog. Phys. 79, 106901 (2016), arXiv:1511.07586 [gr-qc]; H. M. Sadjadi, Phys. Lett. B 718, 270 (2012); S. Carloni, F. S. N. Lobo, G. Otalora, and E. N. Saridakis, Phys. Rev. D 93, 024034 (2016); Y. Zhang, H.  Li, Y.  Gong, and Z. H. Zhu, JCAP, 07, 015 (2011); K. Karami, S. Asadzadeh, A. Abdolmaleki, and Z. Safari, Phys. Rev. D 88, 084034 (2013); M. E. Rodrigues, M. J. S. Houndjo, D. Momeni, and R. Myrzakulov, arXiv:1302.4372v2 [physics.gen-ph]; F. Darabi, M. Mousavi, and K. Atazadeh, Phys. Rev. D 91, 084023 (2015); G. G. L. Nashed,
arXiv:1506.08695 [gr-qc]; X. Fu, P. Wu,  and H. Yu, Int. J. Mod. Phys. D, 21, 09 (2012) 1250074; K. Myrzakulov, P. Tsyba, and R. Myrzakulov, arXiv:1601.07357 [physics.gen-ph];
E. L. B. Junior, M. E. Rodrigues, I. G. Salako, and M. J. S. Houndjo, Class. Quantum Grav. 33, 125006 (2016); A. Behboodi, S. Akhshabi, and K. Nozari, Phys. Lett. B 718, 30 (2012);
M. Sharif and S. Rani, Mod. Phys. Lett. A 28, 1350118 (2013); K. Bamba, S. D. Odintsov, and E. N. Saridakis, arXiv:1605.02461 [gr-qc]; K. Bamba, S. Nojiri and S. D. Odintsov,
Phys. Lett. B 725, 368 (2013); R. C. Nunes, S. Pan, and E. N. Saridakis, arXiv:1606.04359 [gr-qc]
\bibitem{quint}L. Jarv and  A. Toporensky, Phys. Rev. D 93, 024051 (2016); M. Skugoreva and  A. Toporensky, arXiv:1605.01989 [gr-qc]; B. Fazlpour, arXiv:1604.03080 [gr-qc];
C. Q. Geng, C. C. Lee, E.  N. Saridakis, and Y. P. Wu, Phys. Lett. B 704, 384 (2011); G. Otalora, JCAP 07, 044 (2013); J. A. Gu, C. C. Lee, and C. Q. Geng, arXiv:1204.4048v2 [astro-ph.CO];
K. Bamba, S. D. Odintsov and D. Saez-Gomez, Phys. Rev. D 88, 084042 (2013)
\bibitem{sad}H. M. Sadjadi, Phys. Rev. D 87, 064028 (2013)
\bibitem{Wei}Z. C. Chen, Y. Wu, and H. Wei, Nucl. Phys. B 894, 422 (2015)
\bibitem{Li}J. T. Li, Y. P. Wu, and C. Q. Geng, Phys. Rev. D 89, 044040 (2014)
\bibitem{ppn}C. M. Will, Theory and experiment in gravitational physics, Cambridge, UK: Univ. Pr. (1993);  C. M. Will, Liv. Rev. Rel. 9, 3 (2006); L. L. Smalley, Phys. Rev. D21, 328 (1980); M.  Hohmann, Phys. Rev. D 92, 064019 (2015); M.  Hohmann, Phys. Rev. D 92, 064019 (2015); Z. Yi and  Y. Gong, arXiv:1512.05555 [gr-qc]; Kh. Saaidi, A. Mohammadi, and H. Sheikhahmadi, 	Phys. Rev. D 88, 084054 (2013)

\bibitem{thermo}M. Zubair and S. Bahamonde, arXiv:1604.02996 [gr-qc]
\bibitem{jcap} H. Mohseni Sadjadi, JCAP 01, 031 (2017), arXiv:1609.04292 [gr-qc]
\bibitem{boundary}S. Bahamonde, C. G. Boehmer, and  M. Wright, Phys. Rev. D 92, 104042 (2015), arXiv:1508.05120v2 [gr-qc]. M. Wright, Phys. Rev. D 93, 103002 (2016).
\bibitem{cas}B. Bertotti, L. Iess and P. Tortora, Nature 425, 374 (2003)
\bibitem{living}C. M. Will, Living Rev. Rel. 9, 3 (2006), arXiv:gr-qc/0510072, Table 4.
\bibitem{jarv}M. Hohmann, L. Jarv, P. Kuusk, and E. Randla, Phys. Rev. D 88, 084054 (2013)
\bibitem{lunar}F. Hofmann, J. M$\ddot{u}$ller, and L. Biskupek, Astronomy and Astrophysics 522, L5 (2010)
\bibitem{tam}N. Tamanini and C. G. Boehmer, Phys. Rev. D 86, 044009 (2012)
\bibitem{hay}J. Hayward, Phys. Rev. D20, 3039 (1979);  J. Hayward, Gen. Rel. Grav. 13, 43 (1981)


\end{thebibliography}
\end{document}